\newif\ifdoublecol
\doublecoltrue
\ifdoublecol
  \documentclass[twocolumn,fleqn,10pt]{article}
\else
  \documentclass[fleqn,10pt]{wlscirep}
\fi
\usepackage{amsmath,amssymb}
\usepackage{graphicx}
\usepackage{booktabs}
\usepackage{hyperref}
\usepackage{tikz}
\usetikzlibrary{arrows.meta,positioning,calc}
\usepackage{subcaption} 
\usepackage{authblk}
\usepackage[english]{babel}
\usepackage[capitalise]{cleveref}

\newif\ifscirep
\ifdoublecol
  \scirepfalse
\else
  \scireptrue
\fi

\newcommand{\paperabstract}{%
	Parallel p-bit Ising machines provide a promising hardware platform for fast and energy-efficient combinatorial optimization,
	but their scalability and efficiency critically depend on update synchronization, hardware timing, and architectural cost.
	Here we develop a unified performance--cost landscape for parallel p-bit annealing by systematically analyzing synchronous and
	asynchronous update schemes under realistic constraints, including finite hardware delay, time-multiplexed p-bit reuse,
	and limited input digital-to-analog (DAC) precision.
	We show that synchronous updates are not inherently unstable, but can suffer from oscillations when excessive update
	simultaneity is present, while asynchronous updates are structurally constrained by hardware delay and require slower operation
	to maintain stability. To bridge performance and hardware efficiency, we introduce time-multiplexed reuse of physical p-bits
	combined with structured synchronous control policies, which preserve statistically valid annealing dynamics while reducing
	the effective update rate. This reuse decouples statistical correctness from physical resource count, enabling the number of
	physical p-bits and input DACs to scale approximately as the inverse of the time-multiplexing reuse factor.
	As a result, synchronous architectures access low-cost operating regimes, achieving comparable or superior solution quality
	at less than half the normalized hardware cost of optimized asynchronous updates on G-set MaxCut benchmarks with
	800--2000 nodes under matched annealing time. We further demonstrate that low-resolution input DACs (typically 3--4 bits)
	are often sufficient to achieve performance within a few percent of the best-known solutions
	(normalized cut values $\gtrsim 0.95$) when annealing time is appropriately adjusted.
	Together, these results establish coordinated time-multiplexed p-bit reuse combined with structured synchronous control
	as a key architectural principle for scalable probabilistic computing hardware, and provide reproducible design guidance
	for balancing solution quality, hardware cost, and timing constraints under realistic delay and precision limitations.
}

\ifscirep
\begin{abstract}
\paperabstract
\end{abstract}
\fi

\title{A Unified Performance--Cost Landscape of Parallel p-bit Ising Machines Based on Update Dynamics}

\author[*]{Naoya Onizawa}
\author{Takahiro Hanyu}
\affil{Research Institute of Electrical Communication, Tohoku University, Sendai, 980-8577, Japan}
\affil[*]{Corresponding author: naoya.onizawa.a7@tohoku.ac.jp}

\begin{document}
\flushbottom
\maketitle
\ifscirep\else
\begin{abstract}
\paperabstract
\end{abstract}
\fi

\section*{Introduction}
Probabilistic computing is an emerging computing paradigm that exploits intrinsic
stochasticity to efficiently solve problems such as inference, sampling, and
combinatorial optimization~\cite{stochastic_book,chowdhury2023fullstack}.
Its fundamental building block is the \emph{probabilistic bit} (p-bit), a stochastic binary unit whose output fluctuates in time with a tunable mean determined by a weighted sum of inputs~\cite{p-bit_device}.
Networks of interacting p-bits can implement invertible Boolean logic and naturally sample from Boltzmann distributions, enabling hardware-friendly realizations of Ising and QUBO models~\cite{p-bit_general,HardwareEmuSciRep2017,CamsariPRX2017,CIL}.
Compared with fully digital optimization approaches~\cite{SA1}, 
p-bit networks provide a promising route to domain-specific accelerators 
that trade numerical precision for massive parallelism and improved energy efficiency,
particularly when stochastic devices such as low-barrier magnetic tunnel junctions (MTJs)
provide randomness at nanosecond time scales~\cite{p-bit_device_fast1,p-bit_device_fast2}.

Early studies established the principles of probabilistic spin logic, and subsequent device- and circuit-level work demonstrated fast stochastic MTJs and on-chip p-bit cores~\cite{p-bit_device_fast1,p-bit_device_fast2,p-bit_variability}. These developments have motivated applications of p-bit networks to probabilistic inference and optimization, while related hardware approaches to Ising optimization have also been explored in digital annealing, coherent Ising machines, simulated bifurcation, and quantum annealing~\cite{CIL_training,p-bit_BI,p-bit_PT,p-bit_gibbs,p-bit_general,SSA,SSQA,DA,CIM,SB,QA1,QA_review}.

Among these directions, simulated annealing using p-bits (pSA) is attractive because it can be implemented on conventional computing platforms while still exploiting the parallelism inherent to p-bit networks. Unlike traditional simulated annealing, which typically updates nodes sequentially, pSA permits parallel updates, but its performance can degrade on larger problems such as MAX-CUT G-set instances~\cite{G-set,SSA}. Recent algorithmic variants have improved large-scale pSA performance~\cite{TApSA}, which further motivates a systematic study of how update dynamics and hardware constraints influence scalability.

The introduction of these algorithms has opened up the possibility of large-scale implementations of pSA.
However, their effectiveness and scalability depend critically on the underlying update dynamics of interacting p-bits.
At the algorithmic level, the dynamics of p-bit networks during annealing are strongly influenced by how p-bits are \emph{updated}.
%
%
In synchronous, clock-driven designs, many p-bits update simultaneously, which simplifies
control and enables structured memory access, as commonly assumed in parallel Ising
machine implementations~\cite{p-bit_gibbs}.
However, strong simultaneity can induce oscillatory behavior in tightly coupled graphs,
preventing monotonic energy reduction and hindering convergence.
Similar observations have been reported in the context of parallel Monte Carlo
simulations of spin models, where excessive simultaneous updates are known to
introduce strong correlations and degrade mixing, while controlled parallelism
can retain correct sampling behavior~\cite{Weigel2012ParallelMC}.
Asynchronous, event-driven designs naturally desynchronize updates and can mitigate such oscillations, but their behavior depends sensitively on update schedules, communication delays, and the staleness of inputs.
When p-bits act on outdated information, convergence can slow and solution quality can degrade,
a phenomenon closely related to classical results on mixing times and parallel Markov
chain updates~\cite{levin2009markov}.
These effects are closely related to classical sampling theory for Gibbs samplers and their parallel variants~\cite{DeSaPMLR2016,GonzalezChromaticGibbs2011}, as well as recent theoretical analyses showing
that asynchronous updates can introduce bias and slow mixing when operating on
stale information~\cite{terenin2016asyncgibbs,Johnson2013Gibbs,Bertsekas1989Parallel}.

Beyond algorithmic considerations, scalability is fundamentally constrained by hardware resources.
%
%
Implementing weighted interactions among p-bits requires interconnect, memory bandwidth,
and input digital-to-analog conversion (DAC) or equivalent mixed-signal resources, whose
cost can dominate at large problem sizes, as demonstrated in recent experimental p-bit
and Ising machine implementations~\cite{Si2024NatCommun}.
As a result, practical p-bit accelerators must jointly co-design the \emph{update policy} and the \emph{mapping} from a logical p-bit network to physical hardware, balancing solution quality against area, power, and throughput.

This paper develops a unified \emph{performance--cost landscape} for parallel p-bit updates by systematically sweeping four key architectural parameters:
(i) the update policy,
(ii) the update interval $\tau$,
(iii) the time-multiplexing reuse factor $c$, defined as the number of logical p-bits mapped onto a single physical p-bit,
and (iv) the input-DAC bit width $b$.
In particular, we focus on the delay-to-update ratio ($d/\tau$),
which governs the effective coupling between updates and plays a critical role in determining stability,
mixing behavior, and convergence under parallel update schemes.

Our central contribution is a time-multiplexed p-bit reuse scheme ($c>1$) co-designed with synchronous scheduling, which substantially reduces the required number of physical p-bits and DACs while preserving solution quality through appropriate timing and control.

\paragraph{Contributions.}
\begin{itemize}
\item We characterize oscillation and stability regimes for parallel p-bit updates and quantify how convergence depends on the delay-to-update ratio $d/\tau$ and the effective update rate.
\item We propose practical synchronous control strategies (including randomized and structured block schedules) that mitigate harmful simultaneity while preserving hardware-friendly memory access.
\item We introduce a time-multiplexed reuse scheme ($c>1$) and evaluate its performance--cost trade-offs, including reductions in physical p-bits and input-DAC resources relative to one-to-one mappings.
\item We show that low-resolution input DACs (typically 3--4\,bits) can achieve near-best normalized cut when the annealing time is adjusted appropriately, enabling further hardware savings.
\end{itemize}

To address these challenges, this work develops a unified simulation and analysis framework
for studying the architectural trade-offs of parallel p-bit Ising machines.
Rather than proposing a new probabilistic model, the goal is to systematically analyze
how different parallel update policies interact with realistic hardware constraints.

Within this framework, we construct a unified performance--cost landscape by sweeping
four key architectural parameters: update policy, update interval, time-multiplexing
reuse factor, and input-DAC resolution.
All update schemes are evaluated under a common simulation environment with identical
annealing schedules, fixed hardware delay, and matched total simulation time.
This controlled setting allows a fair comparison between synchronous and asynchronous
update mechanisms while isolating the architectural factors that determine scalability,
solution quality, and hardware efficiency.

\begin{figure*}[!t]
    \centering
    \includegraphics[width=0.95\linewidth]{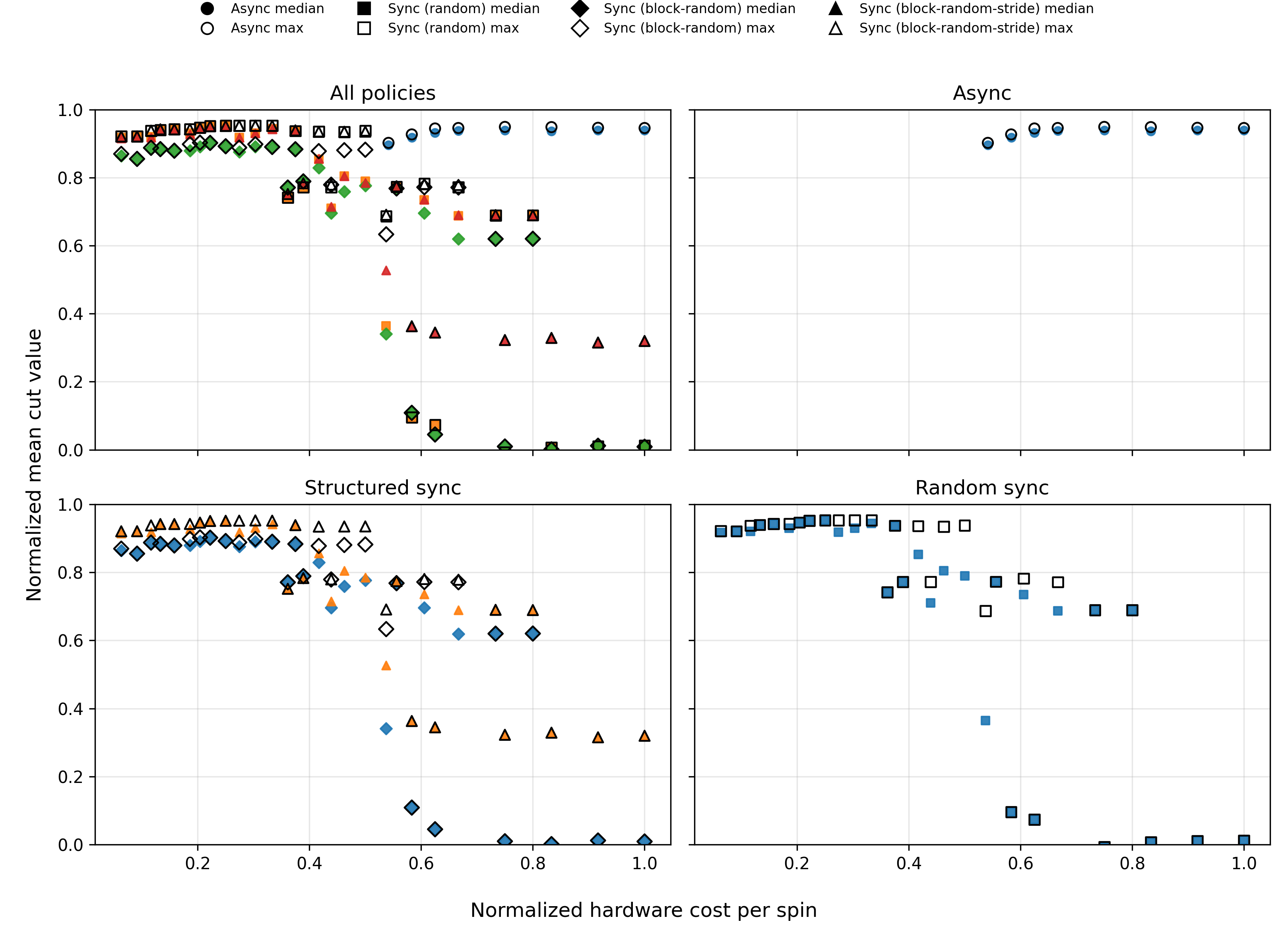}
\caption{
	Unified performance--cost landscape using graph-averaged performance at a fixed simulation time of 500~ns, reorganized into a 2$\times$2 layout for readability.
	The panels show all policies, Async only, structured synchronous schedules, and random synchronous schedules, respectively.
	The horizontal axis shows normalized hardware cost ($C_{\mathrm{HW}}/N$) accounting for p-bit count and DAC resolution, and the vertical axis shows the mean normalized cut value averaged over all G-set instances for each parameter setting.
	Points are binned by hardware cost (40 bins); filled markers indicate the median and open markers the maximum within each bin, and a single global legend is used across all panels.
}
    \label{fig:landscape}
\end{figure*}

\cref{fig:landscape} summarizes the resulting unified performance--cost landscape across different parallel
p-bit update policies on G-set MaxCut benchmarks, highlighting the trade-offs between solution
quality and hardware cost; detailed analyses follow in the Results section.

\section*{Related Work}

\subsection*{p-bit computing and probabilistic spin logic}

Probabilistic bits (p-bits) were originally introduced as stochastic
building blocks for invertible Boolean logic and probabilistic computing
systems~\cite{p-bit_general,CamsariPRX2017}.
Networks of interacting p-bits can emulate Ising models and naturally
perform Boltzmann sampling, enabling hardware realizations of combinatorial
optimization algorithms and probabilistic inference~\cite{HardwareEmuSciRep2017,CIL}.
Experimental and circuit-level studies have demonstrated that stochastic
devices such as low-barrier magnetic tunnel junctions (MTJs) can generate
the required randomness at nanosecond time scales, enabling compact and
energy-efficient hardware implementations~\cite{p-bit_device_fast1,p-bit_device_fast2}.

Beyond combinatorial optimization, p-bit networks have been applied to
a variety of probabilistic tasks, including neural network training,
Bayesian inference, parallel tempering, and Gibbs sampling
\cite{CIL_training,p-bit_BI,p-bit_PT,p-bit_gibbs}.
These results highlight the potential of probabilistic hardware as an
alternative to deterministic digital accelerators for sampling-based
computation.

\subsection*{Update dynamics and parallel sampling}

A central issue in p-bit networks and related stochastic optimization
methods is the update dynamics of interacting variables.
Classical Gibbs sampling theory typically assumes sequential updates,
which guarantee convergence to the correct Boltzmann distribution.
However, sequential updates can limit hardware parallelism.

Parallel or partially synchronous update schemes have therefore been
studied in both statistical physics and machine learning contexts.
For example, parallel Monte Carlo simulations of spin models show that
excessive simultaneous updates can introduce correlations and degrade
mixing behavior~\cite{Weigel2012ParallelMC}.
Similarly, theoretical analyses of parallel Gibbs sampling demonstrate
that asynchronous updates operating on stale information may introduce
bias and slow convergence~\cite{terenin2016asyncgibbs,Johnson2013Gibbs,
	Bertsekas1989Parallel}.

Recent work on p-bit simulated annealing has explored various update
strategies and algorithmic improvements to enable large-scale
optimization~\cite{SSA,TApSA}.
However, the interaction between update policy, hardware timing
constraints, and architectural cost has not been systematically
characterized.

\subsection*{Hardware constraints in Ising machine implementations}

In practical hardware implementations of p-bit or Ising machines,
scalability is often limited not only by algorithmic considerations
but also by architectural resources such as interconnect bandwidth,
memory access, and mixed-signal components.
In particular, computing weighted inputs typically requires
digital-to-analog converters (DACs) or equivalent analog circuitry,
whose area and power can dominate system cost in large-scale
implementations~\cite{Si2024NatCommun}.

These constraints motivate architectural techniques such as
time-multiplexing and resource sharing to reduce the number of
physical computational elements required to emulate a larger
logical network.

\subsection*{Position of this work}

The present work focuses on the architectural implications of
parallel update dynamics in p-bit Ising machines.
Rather than introducing a new probabilistic model,
we develop a unified simulation framework that enables systematic
comparison of synchronous and asynchronous update policies under
explicit hardware constraints.

By jointly analyzing update timing, delay effects, time-multiplexed
p-bit reuse, and DAC precision, we construct a unified
performance--cost landscape that clarifies how architectural
choices influence scalability and solution quality in large-scale
p-bit annealing systems.

\section*{Model and Hardware Assumptions}

This section defines the probabilistic-bit model, timing notation, and abstract hardware
assumptions used throughout this work.
Implementation-specific simulation procedures and update-policy algorithms are described in
the Methods section.

\subsection*{Probabilistic-bit model and Ising formulation}

We consider Ising optimization problems with spins $\sigma_i \in \{\pm 1\}$ and energy
\begin{equation}
	H(\boldsymbol{\sigma}) = -\tfrac{1}{2}\boldsymbol{\sigma}^\top J \boldsymbol{\sigma}
	- \mathbf{h}^\top \boldsymbol{\sigma}.
\end{equation}
Each Ising spin is implemented as a \emph{logical p-bit}, a stochastic binary unit whose output
fluctuates in time depending on its input.
The probabilistic update rule of a p-bit can be expressed as
\begin{equation}
	\sigma_i(t^+) = \mathrm{sgn}\Bigl(r_i(t) + \tanh\bigl(I_i(t)\bigr)\Bigr),
	\label{eq:pbit_update}
\end{equation}
where $r_i(t)\in[-1,1]$ is an independently and uniformly distributed random variable at each
update event, and $I_i(t)$ is a real-valued input signal.
This formulation is equivalent to sampling $\sigma_i(t^+)$ from the Bernoulli distribution
\begin{equation}
	\Pr\bigl[\sigma_i(t^+)=+1\bigr]
	= \tfrac{1}{2}\bigl(1+\tanh I_i(t)\bigr),
\end{equation}
ensuring consistency with Boltzmann statistics.

For Ising optimization, the input to each p-bit is computed as
\begin{equation}
	I_i(t) = I_0(t)\Bigl(h_i + \sum_j J_{ij}\,\sigma_j(t)\Bigr),
\end{equation}
where $h_i$ and $J_{ij}$ denote the bias and coupling weights of the Ising model, respectively.
The parameter $I_0(t)$ acts as a \emph{pseudo inverse temperature} that controls the interaction
strength and is increased over time to implement simulated annealing.

In this work, the p-bit dynamics are modeled at the algorithmic level rather than through a device-level MTJ switching model.
The stochasticity of the updates is controlled through the annealing factor $I_0(t)$.
This factor plays a role analogous to inverse temperature in an Ising system: increasing $I_0(t)$ biases updates more strongly toward lower-energy states.
Accordingly, $I_0(t)$ is varied through the prescribed annealing schedule, while device-level MTJ temperature physics is outside the scope of the present model.

In hardware-oriented implementations, the random signal $r_i(t)$ may be provided by intrinsic
device-level stochasticity (e.g., thermal fluctuations in low-barrier MTJs).
Moreover, the input field $I_i(t)$ may be quantized due to finite input digital-to-analog
converter (DAC) resolution, allowing the impact of limited precision to be explicitly captured.

\subsection*{Time-multiplexing factor $c$ and timing}

\begin{figure*}[!t]
	\centering
	
	\begin{subfigure}[t]{0.48\linewidth}
		\centering
		\begin{tikzpicture}[
			font=\small,
			box/.style={draw, rounded corners, minimum width=1.3cm, minimum height=0.75cm, align=center},
			arrow/.style={-Latex, thick},
			lab/.style={font=\small}
			]
			\node[lab] at (0,2.2) {(a) No time-multiplexing ($c=1$)};
			
			\node[lab, anchor=east] at (-2.6,1.2) {Logical p-bits};
			\node[box] (L1) at (-0.8,1.2) {$L_1$};
			\node[box] (L2) at (0.6,1.2) {$L_2$};
			\node[box] (L3) at (2.0,1.2) {$L_3$};
			
			\node[lab, anchor=east] at (-2.6,-0.2) {Physical p-bits};
			\node[box] (P1) at (-0.8,-0.2) {$P_1$};
			\node[box] (P2) at (0.6,-0.2) {$P_2$};
			\node[box] (P3) at (2.0,-0.2) {$P_3$};
			
			\draw[arrow] (P1.north) -- (L1.south);
			\draw[arrow] (P2.north) -- (L2.south);
			\draw[arrow] (P3.north) -- (L3.south);
			
			\draw[arrow] (-2.0,-1.2) -- (2.4,-1.2);
			\node[lab] at (2.6,-1.2) {$t$};
			
			\node[lab, align=left] at (0,-2.0) {$\bullet$ Dedicated mapping\\[-1mm]
				$\bullet$ Fast updates, higher hardware count};
		\end{tikzpicture}
	\end{subfigure}
	\hfill
	\begin{subfigure}[t]{0.48\linewidth}
		\centering
		\begin{tikzpicture}[
			font=\small,
			slot/.style={draw, rounded corners, minimum width=1.2cm, minimum height=0.7cm, align=center},
			arrow/.style={-Latex, thick},
			lab/.style={font=\small}
			]
			\node[lab] at (0,2.2) {(b) Time-multiplexing ($c>1$)};
			
			\node[lab] at (-0.6,1.55) {$t_0$};
			\node[lab] at (0.8,1.55) {$t_1$};
			\node[lab] at (2.2,1.55) {$t_2$};
			
			\node[lab, anchor=east] at (-2.6,0.8) {Physical $P_1$};
			\node[slot] (P1t0) at (-0.6,0.8) {$L_1$};
			\node[slot] (P1t1) at (0.8,0.8) {$L_3$};
			\node[slot] (P1t2) at (2.2,0.8) {$L_5$};
			\draw[arrow] (P1t0) -- (P1t1);
			\draw[arrow] (P1t1) -- (P1t2);
			
			\node[lab, anchor=east] at (-2.6,-0.4) {Physical $P_2$};
			\node[slot] (P2t0) at (-0.6,-0.4) {$L_2$};
			\node[slot] (P2t1) at (0.8,-0.4) {$L_4$};
			\node[slot] (P2t2) at (2.2,-0.4) {$L_6$};
			\draw[arrow] (P2t0) -- (P2t1);
			\draw[arrow] (P2t1) -- (P2t2);
			
			\draw[arrow] (-2.0,-1.2) -- (2.6,-1.2);
			\node[lab] at (2.8,-1.2) {$t$};
			
			\node[lab, align=left] at (0,-2.05) {$\bullet$ One physical p-bit emulates $c$ logical p-bits\\[-1mm]
				$\bullet$ Effective update interval $\approx c\tau$\\[-1mm]
				$\bullet$ Hardware p-bits/DACs reduced by $\sim 1/c$};
		\end{tikzpicture}
	\end{subfigure}
	
	\caption{Conceptual illustration of time-multiplexed p-bit reuse. (a) Without time-multiplexing ($c=1$), each logical p-bit maps to a dedicated physical p-bit. (b) With time-multiplexing ($c>1$), a physical p-bit sequentially emulates multiple logical p-bits across time slots, reducing physical resources while increasing the effective update interval.}
	\label{fig:time_multiplexing}
\end{figure*}
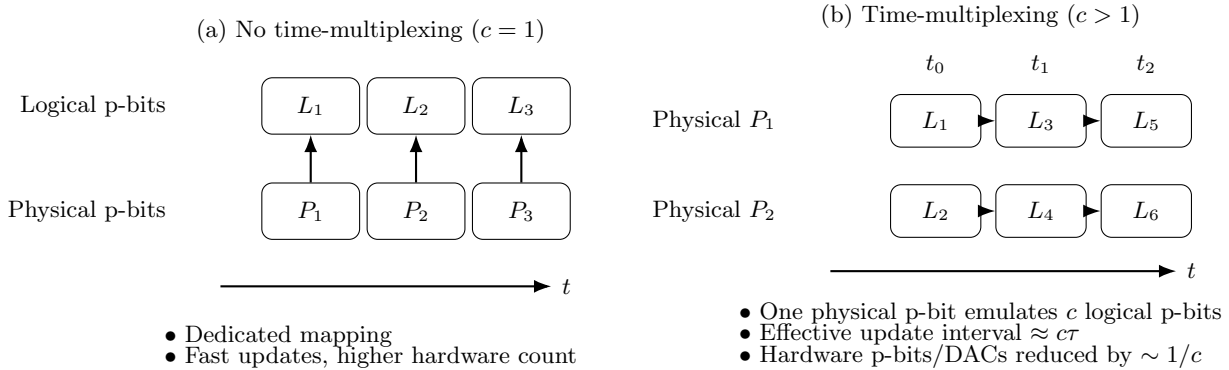
\cref{fig:time_multiplexing} illustrates the concept of time-multiplexed p-bit reuse.
We introduce a \emph{time-multiplexing reuse factor} $c$, defined as the number of logical p-bits
that are sequentially mapped onto a single physical p-bit.
When $c>1$, a single physical p-bit emulates multiple logical spins over successive time slots,
reducing the required number of physical p-bits and input DACs by approximately a factor of $c$.

Let $\tau$ denote the intrinsic update period of a physical p-bit.
Under time-multiplexing, a single physical p-bit sequentially updates $c$ logical spins
over successive time slots, such that the effective flip/update rate per logical spin becomes
\begin{equation}
	\lambda_{\mathrm{spin}} = \frac{1}{\tau c},
\end{equation}
where $\tau$ denotes the intrinsic update period of a physical p-bit.
We assume a fixed apply delay 
$d=5$ ns to capture device/interconnect/control latency. This is consistent with experimental reports of stochastic MTJs exhibiting nanosecond-scale fluctuation dynamics and nanosecond operation, which sets the natural time scale for p-bit updates~\cite{p-bit_device_fast1,p-bit_device_fast2}.
The delay-to-update ratio therefore quantifies the relative severity of delayed updates.

\subsection*{Abstract hardware cost metrics}

We evaluate architectural hardware cost using abstract metrics that capture dominant scaling
behavior rather than technology-specific circuit details.
The dominant cost $C_{\mathrm{HW}}$ accounts for the primary scalable resources: physical p-bits
and input DACs.
With time-multiplexed reuse factor $c$, the number of physical p-bits scales as
\begin{equation}
	N_{\mathrm{p}} = \left\lceil \frac{N}{c} \right\rceil ,
\end{equation}
and we assume $N_{\mathrm{DAC}}\propto N_{\mathrm{p}}$.
We normalize DAC resolution as $\tilde b=b/b_{\mathrm{ref}}$ with $b_{\mathrm{ref}}=12$ \cite{Si2024NatCommun} and define
\begin{equation}
	C_{\mathrm{HW}} = \alpha N_{\mathrm{p}} + \beta\, \tilde b\, N_{\mathrm{DAC}},
	\label{eq:chw}
\end{equation}
where $\alpha$ and $\beta$ are positive weighting factors used only to capture relative scaling.

In addition, parallel architectures incur update-policy-dependent access/control overhead related
to memory access, address generation, random-number sourcing, and verification complexity.
We refer to these effects collectively as $C_{\mathrm{ACC}}$ and summarize their qualitative trends
in \cref{tab:hw_cost}.

\begin{table*}[t]
	\centering
	\caption{
		Qualitative comparison of access and control overhead $C_{\mathrm{ACC}}$ across update policies.
		We summarize dominant architectural implications in terms of memory access patterns,
		address-generation complexity, compatibility with time-multiplexed reuse ($c>1$), and
		verification difficulty.
		Here, ``random source'' refers to a generic source of stochasticity used for spin selection
		or update masking, and ``verification'' refers to relative complexity in functional validation
		and debugging under stochastic control.
	}
	\label{tab:hw_cost}
	\begin{tabular}{lcccc}
		\toprule
		Policy & Random access & Address generation & $c>1$ & Verification \\
		\midrule
		Gillespie (Async.) 
		& high 
		& per-spin random timers 
		& difficult 
		& high \\
		Tick-random (Sync.) 
		& high 
		& random mask generation 
		& easy 
		& moderate \\
		Tick-block-random (Sync.) 
		& low 
		& random start + counter 
		& easy 
		& low \\
		Tick-block-random-stride (Sync.) 
		& low 
		& random start + stride 
		& easy 
		& low--moderate \\
		\bottomrule
	\end{tabular}
\end{table*}

\section*{Results}

\subsection*{Overview: performance--cost landscape}

All results are obtained under the common simulation conditions summarized in \cref{tab:simulation_parameters},
and evaluated on the G-set benchmark instances listed in \cref{tab:gset_benchmarks}.
Using the normalized hardware cost $C_{\mathrm{HW}}$ defined in \cref{eq:chw}, \cref{fig:landscape} establishes a
global performance--cost landscape that captures the dominant architectural trade-offs between
solution quality and hardware efficiency.

Each point in \cref{fig:landscape} corresponds to a representative operating condition defined by the update
policy, time-multiplexing reuse factor $c$, update interval $\tau$, and DAC resolution.
This landscape provides a unifying context for the detailed analyses that follow, allowing
individual results to be interpreted as movements along a common performance--cost frontier
rather than isolated parameter sweeps.

\subsection*{Synchronous updates: oscillation and stabilization}

\begin{figure*}[!t]
  \centering
  \includegraphics[width=\linewidth]{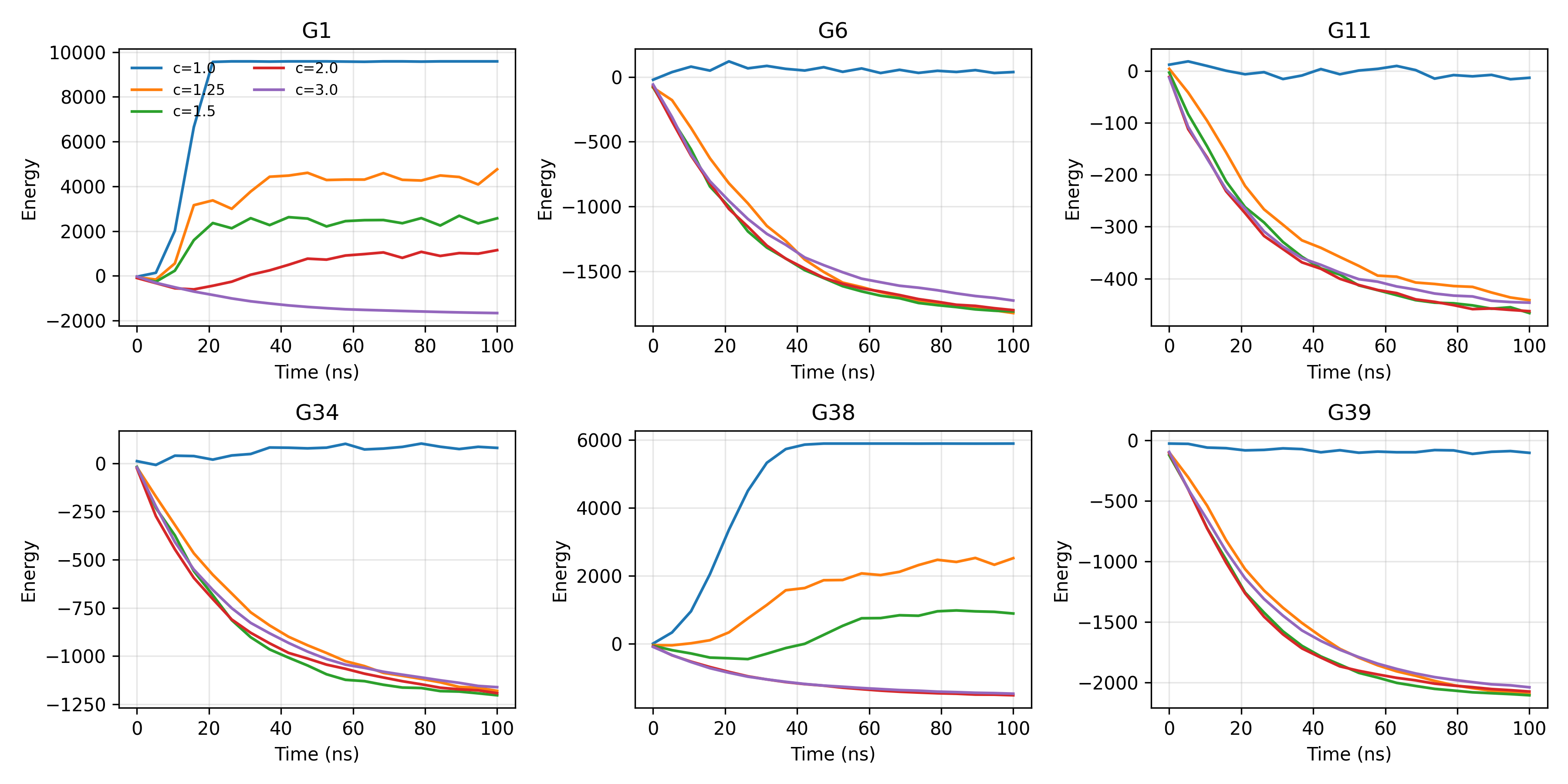}
  \caption{
      Oscillation and stabilization in synchronous random updates.
      Energy time series for six representative G-set instances (G1, G6, G11, G34, G38, and G39)
      under synchronous tick-random updates.
      Each panel compares different time-multiplexing reuse factors
      $c \in \{1, 1.25, 1.5, 2, 3\}$, where larger $c$ corresponds to stronger time-multiplexed reuse
      and a lower effective per-spin update rate.
      For $c=1$, many spins are updated simultaneously at each tick, leading to coherent collective
      switching and pronounced oscillations in the energy trajectory.
      As $c$ increases, update simultaneity is reduced, oscillations are suppressed, and the annealing
      dynamics become progressively more stable across all tested instances.
      A single legend is shown once for the multi-panel figure to avoid repeating the same legend in each panel.
  }
  \label{fig:oscillation}
\end{figure*}

\cref{fig:oscillation} shows representative energy time series under synchronous random updates for six G-set instances
(G1, G6, G11, G34, G38, and G39), with the time-multiplexing reuse factor varied over
$c \in \{1, 1.25, 1.5, 2, 3\}$.
When $c = 1$, a large fraction of logical spins are updated nearly simultaneously at each tick.
In strongly coupled graphs, this simultaneity induces coherent collective switching of many p-bits,
leading to pronounced oscillations in the energy trajectory and preventing effective annealing.

As the time-multiplexing reuse factor $c$ increases, the effective per-spin update rate $\lambda_{\mathrm{spin}} = 1/(\tau c)$
is reduced, and update events become more temporally dispersed.
This dilution of simultaneity breaks phase alignment among spins and suppresses collective oscillations.
For $c \ge 2$, the energy trajectories become markedly smoother across all tested instances,
indicating that annealing dynamics recover stability even under fully synchronous control.

These results highlight that oscillations in synchronous architectures are not an inherent limitation
of clock-driven operation, but rather a consequence of excessive simultaneous updates.
Controlled reduction of the effective update rate—here achieved through time-multiplexed p-bit reuse—
provides a simple and hardware-compatible mechanism for stabilizing synchronous annealing dynamics.

\subsection*{Asynchronous updates: sensitivity to hardware delay}

\begin{figure*}[!t]
  \centering
  \includegraphics[width=\linewidth]{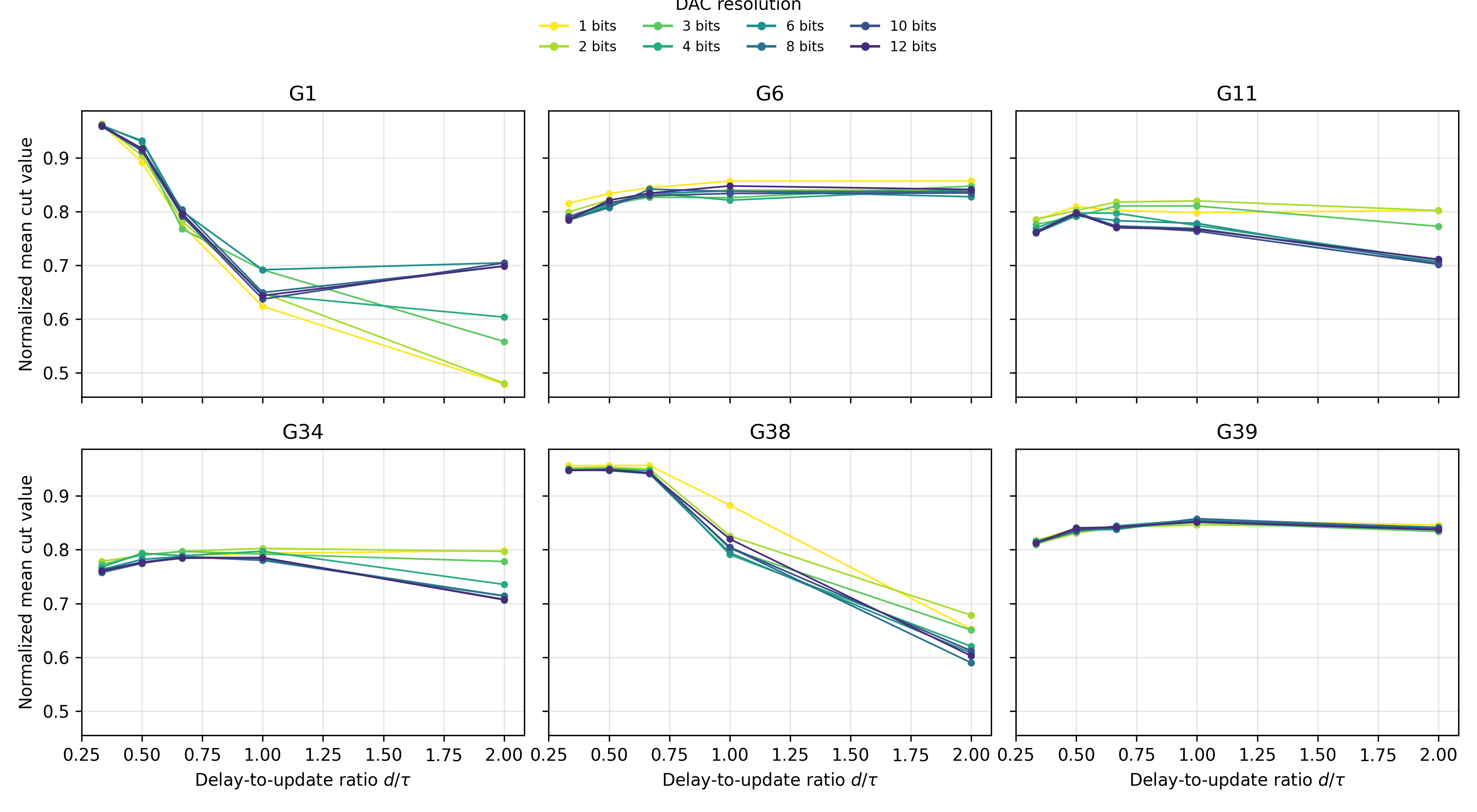}
  \caption{
      Sensitivity of asynchronous (Gillespie-type) updates to hardware delay.
      The normalized mean cut value is plotted as a function of the delay-to-update ratio $d/\tau$ for six
      G-set instances, with the apply delay fixed at $d = 5$~ns and a simulation time of 100~ns.
      Line color indicates the input-DAC resolution, the horizontal axis is labeled explicitly as the delay-to-update ratio $d/\tau$, and a single global legend is used for the full grid.
      For small $d/\tau$, asynchronous updates exhibit stable annealing behavior.
      As $d/\tau$ approaches unity, performance degrades due to spins acting on stale local fields
      computed from outdated neighboring states.
      This effect is exacerbated at low DAC resolution, demonstrating that asynchronous architectures
      are strongly constrained by the relationship between device latency and update interval.
  }
  \label{fig:delay}
\end{figure*}

\cref{fig:delay} quantifies the impact of hardware delay on asynchronous (Gillespie-type) updates by plotting
the normalized mean cut value as a function of the delay-to-update ratio $d/\tau$ for six representative
G-set instances, with the apply delay fixed at $d = 5$~ns and a simulation time of 100~ns.
In the idealized limit $d/\tau \ll 1$, asynchronous updates avoid global simultaneity and exhibit
stable annealing behavior across a wide range of DAC resolutions.

However, as $d/\tau$ approaches unity, performance degrades rapidly.
In this regime, spins increasingly act on stale local fields computed from outdated neighboring states~\cite{Niu2011Hogwild}.
This temporal inconsistency effectively reintroduces correlated update errors, leading to bias and
instability despite the absence of explicit synchronization.
The degradation is particularly pronounced at lower DAC resolutions, where quantization noise
further amplifies the effect of delayed information.

These results demonstrate that asynchrony alone does not guarantee robustness.
Correct and efficient asynchronous annealing requires that the hardware delay remain sufficiently
small relative to the update interval, imposing a nontrivial constraint on clock frequency and
device latency in large-scale implementations.

Although asynchronous updates avoid explicit synchronization, their performance is
highly sensitive to the delay-to-update ratio $d/\tau$, imposing a practical constraint
on clock frequency and preventing access to low-cost operating regimes.

\subsection*{Control policies for synchronous updates}

\begin{figure*}[!t]
  \centering
  \includegraphics[width=\linewidth]{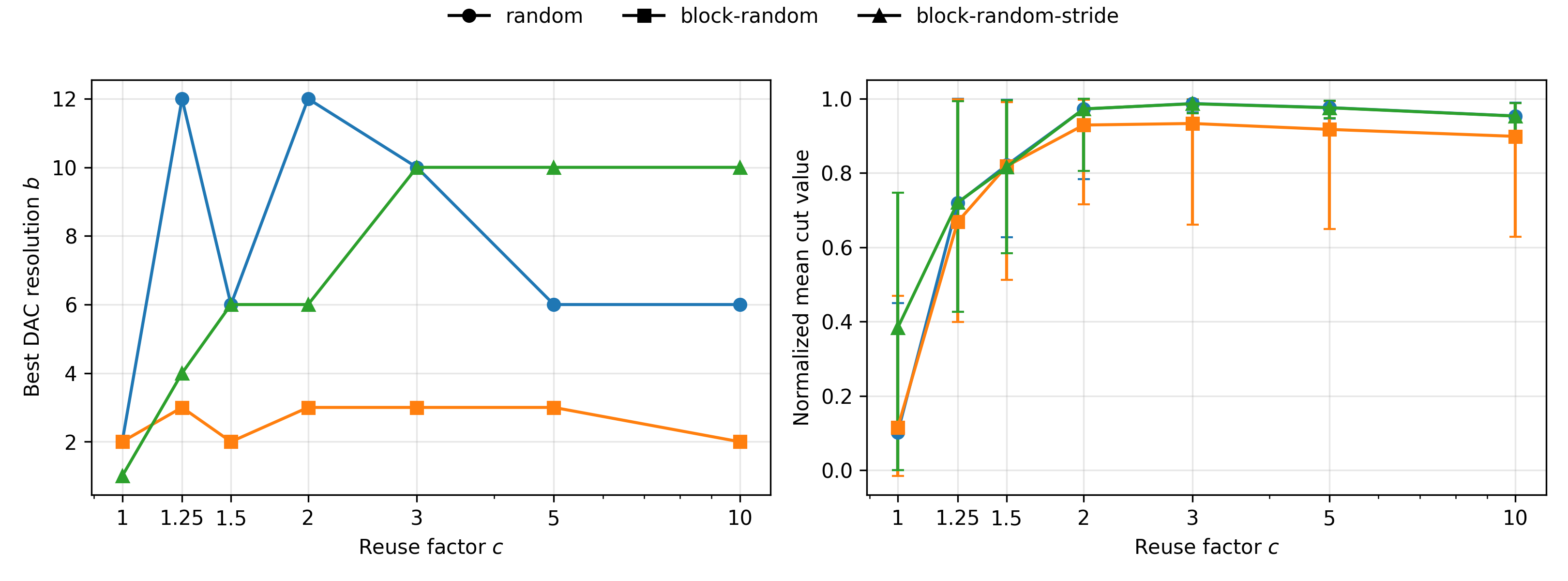}
  \caption{
      Comparison of synchronous update policies.
      Left: the DAC bit width $b$ that maximizes the mean normalized cut value for each reuse factor $c$ and update policy.
      Right: the corresponding mean normalized cut value at that optimal bit width, with error bars indicating the min–max range across G-set instances.
      All results use a simulation time of 500~ns.
      Results are shown for tick-random, block-random, and block-random-stride scheduling policies; block-based schedules attain similar or higher performance with comparable or lower bit width than fully random updates.
  }
   \label{fig:policy}
\end{figure*}

\cref{fig:policy} compares three synchronous control policies—random, block-random, and
block-random-stride—by reporting, for each time-multiplexing reuse factor $c$, the DAC bit width that maximizes
the normalized mean cut value (left) and the corresponding performance at that optimal bit width (right),
using a simulation time of 100~ns. The normalized mean cut value is averaged over all G-set benchmarks
listed in \cref{tab:gset_benchmarks}.

Across all values of $c$, block-based policies consistently achieve comparable or higher performance
than fully random masking while requiring equal or lower optimal DAC resolution.
This behavior indicates that structured randomization suppresses harmful update correlations
without sacrificing effective exploration of the energy landscape.
In particular, block-random-stride combines spatial dispersion of updates with simple address generation,
yielding robust performance over a wide range of compression factors.

From a hardware perspective, these results are significant because block-based policies preserve
contiguous or pseudo-contiguous memory access patterns.
Thus, they simultaneously reduce access and control overhead while maintaining annealing quality,
making them well suited for scalable synchronous p-bit architectures with time-multiplexed reuse.

These results indicate that structured randomness, rather than fully independent
random masking, is sufficient to suppress harmful update correlations while preserving
effective exploration of the energy landscape.

\subsection*{DAC precision and annealing time}

\begin{figure*}[!t]
  \centering
  \includegraphics[width=\linewidth]{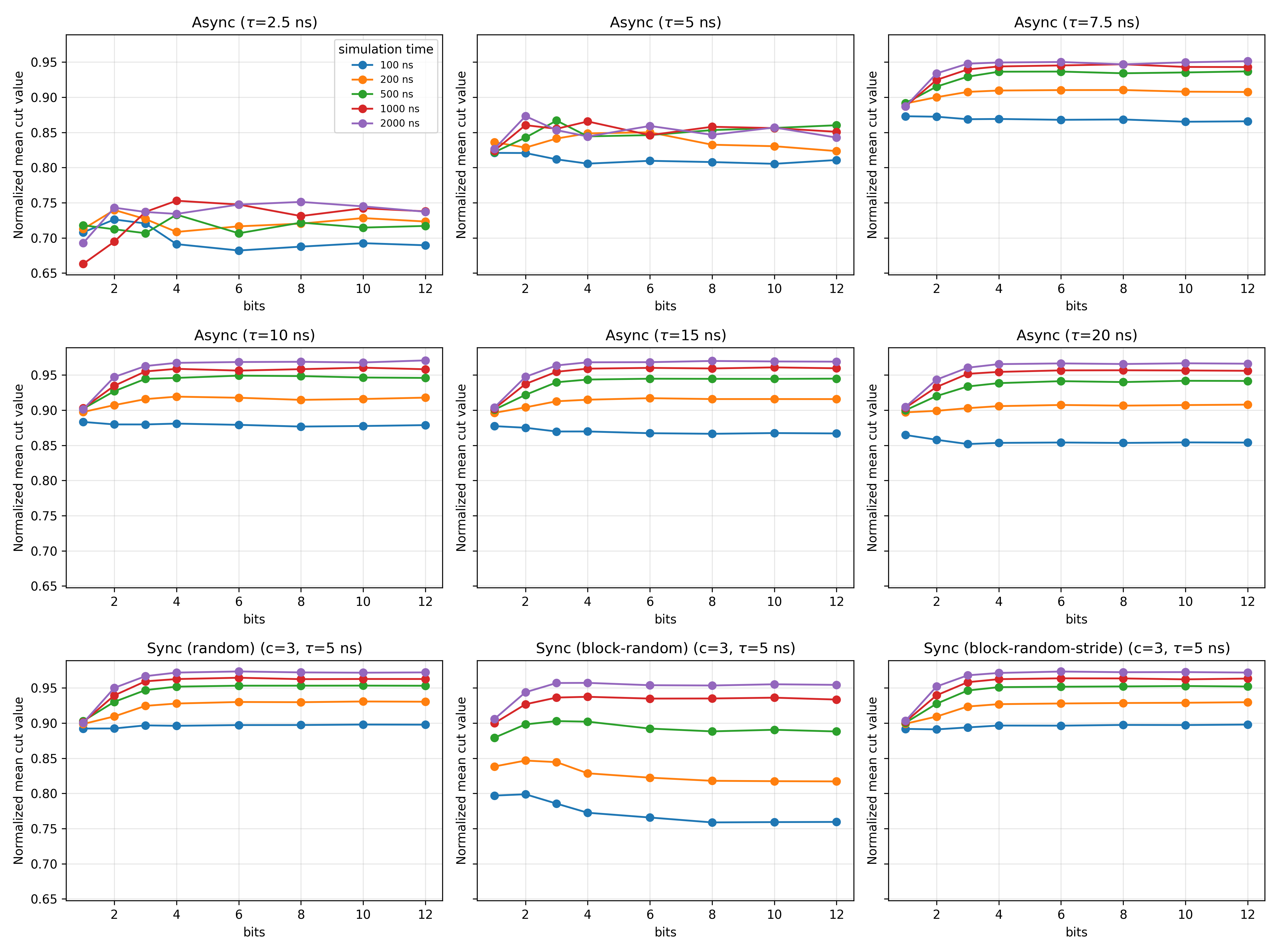}
  \caption{
      Normalized mean cut value as a function of input-DAC resolution for multiple annealing times.
      Top row: asynchronous (Gillespie) updates evaluated at $\tau = 2.5$, 5, 7.5, 10, 15, and 20~ns.
      Bottom row: synchronous updates (random, block-random, and block-random-stride) evaluated at
      $\tau = 5$~ns with time-multiplexing reuse factor $c = 3$.
      Each curve corresponds to a different total simulation (annealing) time.
      Reducing DAC resolution degrades performance when annealing time is fixed, but longer annealing
      partially compensates for coarse quantization.
      Synchronous policies exhibit smoother degradation and greater tolerance to low-resolution DACs
      than asynchronous updates.
  }
  \label{fig:dac}
\end{figure*}

\cref{fig:dac} summarizes the dependence of annealing performance on input-DAC resolution for both
asynchronous and synchronous update schemes.
The normalized mean cut value is plotted as a function of DAC bit width $b$ for multiple annealing
times, with asynchronous updates evaluated at $\tau \in \{2.5, 5, 7.5, 10, 15, 20\}$~ns and synchronous policies
evaluated at $\tau = 5$~ns with $c = 3$.
The normalized mean cut value is averaged over all G-set benchmarks listed in \cref{tab:gset_benchmarks}.

For both update styles, reducing DAC resolution degrades performance when annealing time is fixed.
However, longer annealing partially compensates for coarse quantization by allowing stochastic
averaging over a larger number of update events.
Notably, synchronous policies with controlled update scheduling exhibit a smoother degradation
curve than asynchronous updates, indicating greater tolerance to low-resolution input biases.

These results suggest that DAC precision can be treated as a flexible design parameter rather than
a hard constraint.
In many cases, 3--4 bit DACs are sufficient to achieve near-optimal performance when combined with
moderate increases in annealing time, enabling substantial reductions in hardware area and power
without sacrificing solution quality.

\subsection*{Sequential baseline as an algorithmic reference}

\begin{figure*}[!t]
  \centering
  \includegraphics[width=0.72\linewidth]{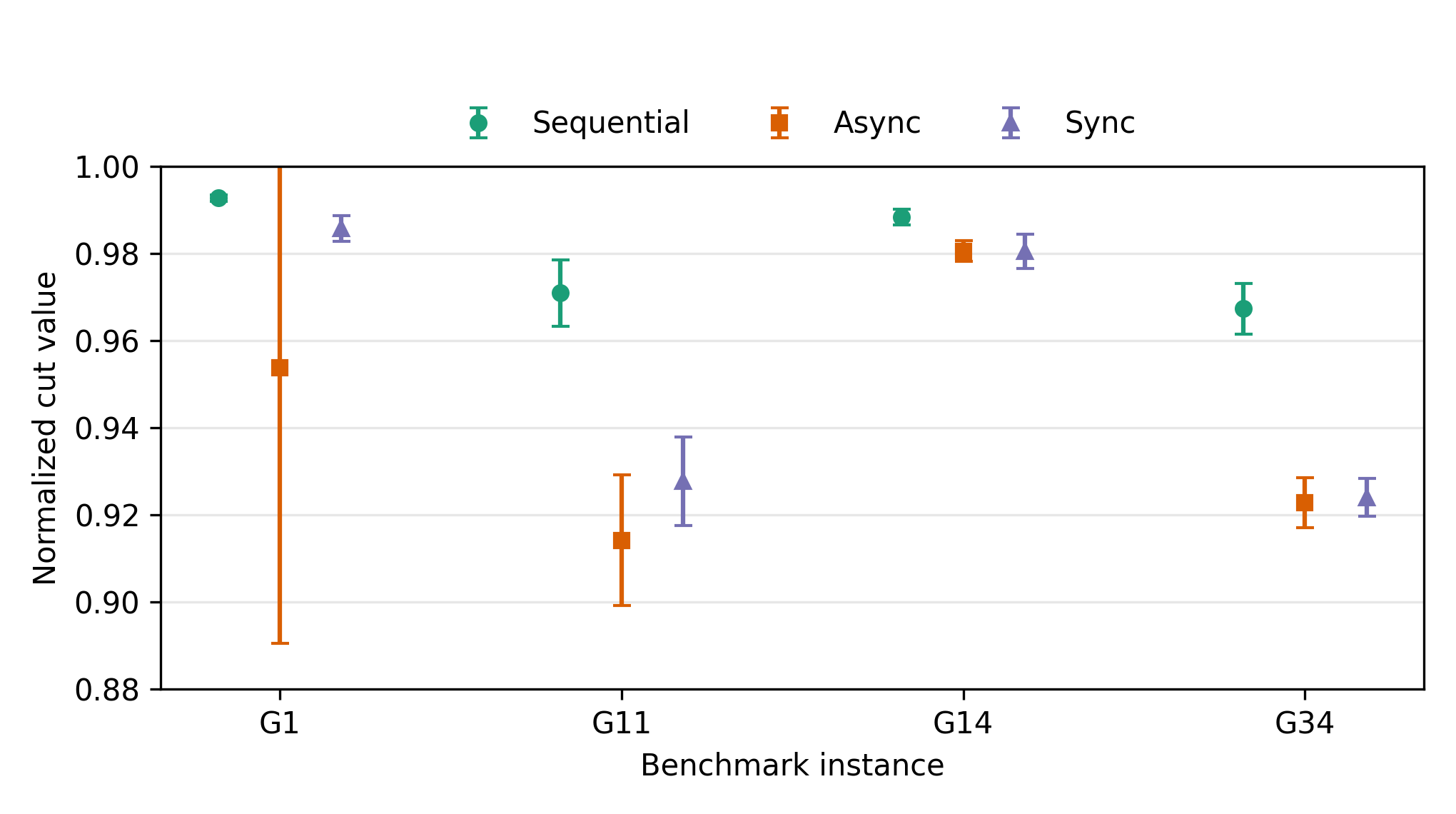}
  \caption{
      Small multi-instance comparison of representative sequential, asynchronous, and synchronous settings on G1, G11, G14, and G34.
      The plotted values are normalized cut values obtained with $b=10$ under a common linear annealing schedule and total simulation time of 500~ns.
      Async denotes Gillespie updates with $\tau=10$~ns and $c=1$, Sync denotes tick block-random-stride updates with $\tau=5$~ns and $c=3$, and Sequential denotes the systematic sequential single-spin update baseline with $\tau=5$~ns and $c=1$.
      Error bars indicate the standard deviation over 5 repeats, and markers are shown without line connections because the benchmark instances are categorical.
      }
  \label{fig:seq_baseline_compare}
\end{figure*}

To provide a compact algorithmic reference against the representative parallel operating points, we additionally compare a sequential baseline against asynchronous and synchronous updates on four G-set instances (G1, G11, G14, and G34), as summarized in \cref{fig:seq_baseline_compare}.
The asynchronous setting uses Gillespie updates with $\tau=10$~ns and $c=1$, the synchronous setting uses tick block-random-stride updates with $\tau=5$~ns and $c=3$, and the sequential baseline uses tick sequential updates with $\tau=5$~ns and $c=1$; all three use $b=10$, the same linear annealing schedule, a total simulation time of 500~ns, and 5 repeats per instance.
Across these tested instances, the sequential baseline attains the highest normalized cut value, while the synchronous and asynchronous representative settings are broadly comparable, with the synchronous setting slightly higher on some instances.
The error bars in \cref{fig:seq_baseline_compare} indicate the standard deviation over the 5 repeated runs for each instance and mode.
Using the four instance-level normalized cut values as paired observations, a Wilcoxon signed-rank test did not reach conventional significance thresholds for any pairwise comparison among the Sequential, Async, and Sync settings (all $p \ge 0.068$), so we report this benchmark as a variability-aware reference comparison rather than as a statistical superiority claim.
Because this benchmark covers only a small set of instances, the sequential mode is used here only as an algorithmic reference rather than as a practical hardware timing model.
In practical hardware implementations, strictly serialized updates would require extremely high clock frequencies for large systems, because all spins would need to be updated sequentially within the same algorithmic time step. Therefore, the sequential mode should be interpreted only as an algorithmic reference rather than a realistic hardware timing model.

\subsection*{Representative operating points under hardware cost constraints}

Representative operating points under explicit hardware cost constraints are summarized in
\cref{tab:rep_operating_points}, using the normalized hardware cost
$C_{\mathrm{HW}}$ defined in \cref{eq:chw}.
All results are obtained at a fixed total simulation (annealing) time of 500~ns and are averaged
over all G-set benchmarks listed in \cref{tab:gset_benchmarks}, ensuring a fair comparison across update policies
without relying on aggressive or policy-dependent timing assumptions.

For asynchronous (Gillespie-type) updates, \cref{tab:rep_operating_points} shows that increasing
the update interval $\tau$ from 5~ns to 15--20~ns significantly improves the normalized mean cut
value.
This confirms that slower updates alleviate the detrimental effects of hardware delay by reducing
the fraction of updates applied using stale information.
However, even at the most favorable update interval considered here, asynchronous updates remain
confined to relatively high normalized hardware cost, as time-multiplexed p-bit reuse ($c>1$) is
not readily supported.
Consequently, performance recovery in asynchronous schemes is achieved primarily through reduced
update rates rather than through improved hardware efficiency.

In contrast, synchronous architectures achieve comparable or higher normalized performance while
operating at substantially lower hardware cost.
As shown in \cref{tab:rep_operating_points}, block-random and block-random-stride scheduling with
time-multiplexed reuse ($c=3$) attain near-best normalized cut values at less than half the
normalized cost of the best asynchronous configurations.
These results indicate that the advantage of synchronous architectures arises not from optimistic
timing assumptions, but from their ability to exploit coordinated reuse of physical p-bits under
explicit hardware cost constraints.

\begin{figure*}[t]
  \centering
  \includegraphics[width=0.9\linewidth]{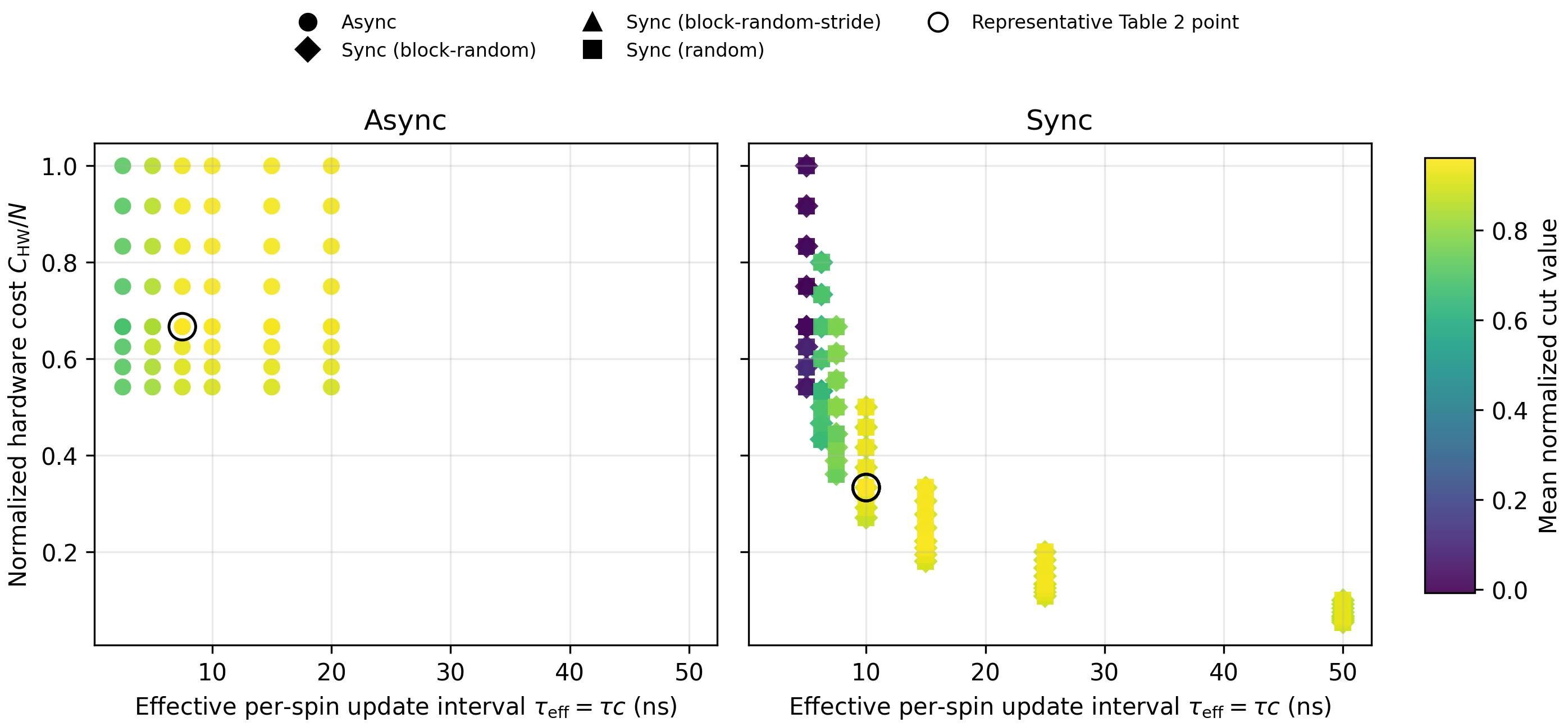}
  \caption{
      Cost--latency visualization derived from the raw 500~ns results underlying \cref{tab:rep_operating_points}.
      Left: asynchronous (Gillespie) configurations. Right: synchronous configurations for random, block-random, and block-random-stride scheduling.
      The horizontal axis shows the effective per-spin update interval $\tau_{\mathrm{eff}}=\tau c$, the vertical axis shows the normalized hardware cost, and color indicates the mean normalized cut value averaged over the same G-set instances used for Table~2.
      Open black circles denote the representative operating points selected in \cref{tab:rep_operating_points}.
      }
  \label{fig:table2_tradeoff}
\end{figure*}

To make the trade-off underlying \cref{tab:rep_operating_points} more explicit, \cref{fig:table2_tradeoff} visualizes the same 500~ns raw sweep data in the cost--latency plane.
Here the horizontal axis is the effective per-spin update interval $\tau_{\mathrm{eff}}=\tau c$, which reduces to $\tau$ for asynchronous updates, and the vertical axis is the normalized hardware cost.
The representative operating points reported in \cref{tab:rep_operating_points} are outlined, showing that synchronous schedules reach lower-cost regions through time-multiplexed reuse, while asynchronous settings improve performance mainly by moving toward larger effective update intervals.

While \cref{tab:rep_operating_points} highlights the best-performing configurations without
restricting DAC resolution, practical hardware designs often impose additional constraints on
mixed-signal precision.
To address this, \cref{tab:lowbit_operating_points} reports the best representative operating
points under a strict low-precision constraint of $b \leq 4$ bits.
Even under this limitation, synchronous update schemes---especially block-random-stride
scheduling---maintain performance close to the unconstrained optimum while preserving the same
low-cost operating regime enabled by time-multiplexed reuse.
Asynchronous updates, by contrast, continue to require higher normalized hardware cost to achieve
similar performance, reflecting their limited tolerance to low DAC resolution and lack of
efficient reuse.

Taken together, \cref{tab:rep_operating_points,tab:lowbit_operating_points} demonstrate that
synchronous architectures with structured scheduling not only dominate the performance--cost
trade-off in the unconstrained case, but also retain their advantage under realistic hardware
constraints on DAC precision.
This robustness underscores the central role of coordinated p-bit reuse and structured synchronous
control in accessing low-cost operating regimes that are fundamentally difficult to reach with
fully asynchronous updates.

\begin{table*}[!t]
    \centering
\caption{
	Best normalized mean cut per update policy averaged across all G-set instances at a fixed
	simulation time of 500~ns.
	Synchronous results are shown for block-random, block-random-stride, and random schedules with
	time-multiplexed reuse ($c>1$), and the row ``Async'' reports the single best asynchronous
	configuration over the full $\tau$ sweep.
	The additional asynchronous rows list the best configurations at fixed update intervals
	$\tau\in\{5,7.5,15,20\}$~ns for comparison, highlighting the sensitivity of Gillespie updates to
	the delay-to-update ratio.
}
\label{tab:rep_operating_points}
    \begin{tabular}{lrrrrr}
    	\toprule
    	Policy & Cut$_{\mathrm{norm}}$ & Cost$_{\mathrm{norm}}$ & $b$ & $c$ & $\tau$~(ns) \\
    	\midrule
    	Async & 0.9492 & 0.7500 & 6 & 1 & 10 \\
    	Sync (block-random) & 0.9029 & 0.2083 & 3 & 3 & 5 \\
    	Sync (block-random-stride) & 0.9528 & 0.3056 & 10 & 3 & 5 \\
    	Sync (random) & 0.9534 & 0.3056 & 10 & 3 & 5 \\
    	\hline
    	Async ($\tau$=5~ns) & 0.8669 & 0.6250 & 3 & 1 & 5 \\
    	Async ($\tau$=7.5~ns) & 0.9368 & 1.0000 & 12 & 1 & 7.5 \\
    	Async ($\tau$=15~ns) & 0.9449 & 1.0000 & 12 & 1 & 15 \\
    	Async ($\tau$=20~ns) & 0.9418 & 0.9167 & 10 & 1 & 20 \\
    	\bottomrule
    \end{tabular}
\end{table*}

\begin{table*}[!t]
	\centering
	\caption{Best mean normalized cut per policy at 500~ns with $b\leq 4$.}
	\label{tab:lowbit_operating_points}
	\begin{tabular}{lrrrr}
		\toprule
		Policy & Cut$_{\mathrm{norm}}$ & Cost$_{\mathrm{norm}}$ & $c$ & $\tau$~(ns) \\
		\midrule
		Async & 0.9460 & 0.6667 & 1 & 10 \\
		Sync (block-random) & 0.9021 & 0.2222 & 3 & 5 \\
		Sync (block-random-stride) & 0.9512 & 0.2222 & 3 & 5 \\
		Sync (random) & 0.9519 & 0.2222 & 3 & 5 \\
		\bottomrule
	\end{tabular}
\end{table*}


\section*{Discussion}

\subsection*{Synchronous versus asynchronous architectures}

The results presented in this work clarify a fundamental architectural trade-off between
synchronous and asynchronous parallel p-bit systems.
Synchronous updates offer explicit global coordination and predictable timing, which are
naturally compatible with hardware scheduling and memory access, but naive clock-driven
operation can induce collective oscillations when many strongly coupled p-bits are updated
simultaneously.
Asynchronous updates, in contrast, avoid explicit synchronization and naturally desynchronize
spin updates, but their stability critically depends on the relationship between the hardware
delay $d$ and the update interval $\tau$.

As demonstrated in the Results section, particularly in \cref{fig:delay} and
\cref{tab:rep_operating_points}, asynchronous annealing degrades rapidly as the ratio $d/\tau$
approaches unity, due to spins acting on stale local fields.
Although this degradation can be partially mitigated by increasing $\tau$, doing so reduces
the effective update rate rather than improving hardware efficiency.
As a consequence, asynchronous architectures remain confined to relatively high hardware cost
regimes and cannot exploit coordinated time-multiplexed reuse of physical p-bits.

These observations indicate that asynchrony should not be regarded as inherently more robust.
Instead, both synchronous and asynchronous schemes are subject to stability constraints that
arise from the interaction between update statistics and hardware timing.
In synchronous systems, excessive simultaneity leads to oscillations, whereas in asynchronous
systems, excessive delay leads to staleness.
This symmetry highlights that stability is governed not by the presence or absence of a clock,
but by the effective update rate relative to hardware latency.

\subsection*{Role of time-multiplexed p-bit reuse}

A central contribution of this work is the demonstration that time-multiplexed reuse of
physical p-bits ($c>1$) enables substantial reductions in hardware cost without altering the
target stationary distribution of the stochastic dynamics.
Importantly, this reuse scheme does not modify the update probabilities of individual logical
spins; rather, it uniformly rescales their effective update rate as
$\lambda_{\mathrm{spin}} = 1/(\tau c)$.
As a result, time-multiplexed reuse corresponds to a temporal rescaling of the underlying
Markov process rather than a change in its transition kernel.

This separation between statistical correctness and temporal efficiency explains why reuse does
not introduce systematic bias.
Each logical spin experiences updates drawn from the same Bernoulli (synchronous) or Poisson
(asynchronous) statistics as in the $c=1$ case, but with a reduced rate.
Such time-thinning arguments are well established in stochastic simulation theory and imply
that only the convergence speed, not the stationary distribution, is affected.

Synchronous architectures are particularly well suited to this approach because they provide
explicit control over update scheduling and naturally align with the discrete time structure
required for reuse.
Fully asynchronous architectures, by contrast, lack a straightforward mechanism to guarantee
statistically consistent reuse across logical spins without introducing additional buffering,
scheduling, and control overhead.
This limitation is structural rather than algorithmic and explains why asynchronous updates
cannot access the low-cost operating regimes observed in \cref{fig:landscape} and summarized in
\cref{tab:rep_operating_points,tab:lowbit_operating_points}.

\subsection*{Structured synchronous control and hardware implications}

The Results further demonstrate that structured synchronous update policies play a critical
role in enabling stable and efficient reuse.
Block-random and block-random-stride scheduling suppress harmful correlations associated with
simultaneous updates while preserving contiguous or pseudo-contiguous memory access patterns.
This combination is particularly advantageous for hardware implementation, as it reduces
address-generation complexity, random-number usage, and memory-access overhead.

From a design perspective, the time-multiplexing reuse factor $c$ and the input-DAC resolution
$b$ emerge as interacting architectural knobs.
Increasing $c$ reduces the number of physical p-bits and DACs approximately as $1/c$, while
reducing $b$ lowers mixed-signal area and power at the cost of increased quantization noise.
The Results show that moderate increases in annealing time can often compensate for both effects,
allowing designers to trade temporal efficiency for substantial hardware savings.

Recent circuit- and device-level advances have further demonstrated that the mixed-signal overhead associated with input digital-to-analog converters can be significantly reduced or even eliminated. 
In particular, DAC-free p-bit architectures based on stochastic nanodevices have been proposed, where probabilistic tunability and annealing are realized entirely through digital delay-based control without explicit analog inputs~\cite{SelcukIEDM2025DACFree}.
From the perspective of the unified performance--cost landscape developed in this work, such DAC-free designs can be interpreted as operating in the extreme low-precision limit of the input representation. In this regime, the dominant architectural trade-offs shift from input resolution toward update dynamics, timing control, and effective update rates.
Accordingly, such DAC-free implementations are most naturally suited to probabilistic inference and combinatorial optimization tasks that tolerate low input precision and timing variability, rather than applications requiring fine-grained numerical control or strict deterministic timing.
Our results therefore provide a complementary, system-level framework for understanding how these emerging DAC-free p-bit implementations may be positioned within a broader design space, clarifying the conditions under which synchronous or asynchronous update strategies can effectively exploit reduced input precision to access low-cost operating regimes under realistic hardware delay constraints.

The representative operating points summarized in \cref{tab:rep_operating_points} demonstrate
that synchronous architectures with structured scheduling dominate the performance--cost
landscape even without constraining DAC precision.
More importantly, \cref{tab:lowbit_operating_points} shows that this advantage persists under
strict low-precision constraints ($b \leq 4$), confirming that the proposed design principles
remain valid under realistic hardware limitations.

\subsection*{Design implications and limitations}

Taken together, these results establish coordinated time-multiplexed reuse combined with
structured synchronous control as a key architectural principle for scalable probabilistic
computing hardware.
The advantage of synchronous architectures does not stem from idealized timing assumptions,
but from their ability to explicitly coordinate reuse and scheduling under finite hardware
delay and limited precision.

Several limitations point to directions for future work.
First, the hardware cost model employed here is intentionally abstract and does not capture
technology-specific circuit details such as DAC area, noise, or power consumption.
Second, device-level variability beyond stochastic bit flips is not explicitly modeled and may
further interact with reuse and scheduling strategies.
Finally, extremely large-scale systems may introduce additional constraints related to memory
bandwidth and interconnect scaling that are not captured by the present model.

Despite these limitations, the unified performance--cost framework developed in this work
provides a systematic basis for evaluating architectural trade-offs in parallel p-bit systems.
By explicitly linking update statistics, hardware timing, and resource reuse, it offers concrete
design guidance for the development of large-scale, energy-efficient probabilistic computing
accelerators.

\section*{Methods}

This section describes the simulation framework, update-policy implementations, and benchmark
settings used throughout this study.
All methodological choices are designed to ensure a fair, reproducible comparison across update
policies, time-multiplexing reuse factors, and hardware constraints.
The simulation parameters and sweep ranges are summarized in \cref{tab:simulation_parameters},
and the benchmark instances are listed in \cref{tab:gset_benchmarks}.

\subsection*{Simulation framework and timing}

All simulations are performed using a custom Python-based framework that implements both
synchronous and asynchronous p-bit update schemes, including explicit modeling of apply delay
and time-multiplexed reuse.
Each logical spin $\sigma_i\in\{\pm1\}$ is updated according to the probabilistic rule defined in
Eqs.~(2)--(4).

For synchronous schemes, time is discretized into global ticks of duration $\Delta t$.
In all synchronous simulations, we choose a hardware-oriented setting
\[
\Delta t=\tau=d,
\]
corresponding to a fully clocked design in which each physical p-bit is updated once per tick and
the updated state is applied at the subsequent tick.

For asynchronous schemes, updates occur in continuous time using a Gillespie-type event-driven
procedure~\cite{Gillespie1977}.
Each logical spin generates update events with effective rate
\[
\lambda_{\mathrm{spin}}=\frac{1}{\tau c},
\]
and proposed updates are applied after a fixed delay $d$.

\subsection*{Update-policy implementations}

We evaluate one asynchronous and three synchronous update policies.

In addition, we use a sequential baseline as an algorithmic reference.
In this mode, spins are updated one-by-one, and each newly updated state is immediately used to compute the local field for the next spin, so that one sweep over a system of $N$ spins corresponds to $N$ sequential single-spin updates.
This baseline is included to provide a standard sequential Gibbs-like reference for comparison with parallel update policies.
A strictly serialized hardware implementation of this procedure would require extremely high clock rates for large $N$, so it should not be interpreted as a practical hardware timing model, but rather as a conceptual algorithmic reference point.

\paragraph{Gillespie (asynchronous).}
Update events are generated in continuous time.
At each event, a spin is selected and updated probabilistically based on its local field; the
resulting state change is applied after delay $d$.
For asynchronous updates, the total simulation (annealing) time is matched to that of synchronous
schemes.
In particular, comparisons are performed at identical annealing durations rather than equal
numbers of update events, avoiding artificial advantages due to higher effective event rates.

\paragraph{Tick-random (synchronous).}
At each tick, an independent Bernoulli mask is generated with probability
$p_{\mathrm{flip}}=\Delta t/(\tau c)=1/c$ for each spin.
All selected spins are updated in parallel and applied after delay $d$.

\paragraph{Tick-block-random (synchronous, contiguous).}
Instead of generating a full random mask, a contiguous block of spins is selected at each tick.
The block starting index is chosen uniformly at random, and the block length is set to
\[
u=\left\lceil p_{\mathrm{flip}}\,N \right\rceil = \left\lceil \frac{N}{c} \right\rceil .
\]
The updated indices are $\mathcal{I}=\{(s+j)\bmod N\}_{j=0}^{u-1}$.

\paragraph{Tick-block-random-stride (synchronous, pseudo-contiguous).}
This variant extends the block-random scheme by introducing a random stride $r$ biased to be
coprime with $N$.
The updated indices are $\mathcal{I}=\{(s+jr)\bmod N\}_{j=0}^{u-1}$, which disperses spatial
correlations while retaining simple and hardware-friendly address generation.

\subsection*{Annealing schedule and quantization}

The pseudo inverse temperature $I_0(t)$ is increased linearly from $I_{0,\min}$ to $I_{0,\max}$
over the total simulation time.
We use
\[
I_{0,\min} = \frac{0.1}{\sigma}, \qquad
I_{0,\max} = \frac{10}{\sigma},
\]
where $\sigma = \sqrt{(N-1)\,\mathrm{Var}(J_{i,:})}$ normalizes coupling statistics across problem
instances~\cite{TApSA}.

Input fields are quantized using a $b$-bit DAC model.
The DAC resolution $b$ is swept over the range listed in \cref{tab:simulation_parameters}.

\subsection*{Benchmarks and performance metrics}

\begin{table*}[t]
	\centering
	\caption{G-set MaxCut benchmark instances used in this work.
		$N$ is the number of vertices, $M$ is the number of edges, and ``Target'' denotes the
		best-known cut value reported in the G-set benchmark suite.}
	\label{tab:gset_benchmarks}
	\begin{tabular}{lrrrrl}
		\hline
		Graph & $N$ & $M$ & Weight & Type & Target \\
		\hline
		G1  & 800  & 19176 & +1       & random   & 11624 \\
		G6  & 800  & 19176 & +1, --1  & random   & 2178  \\
		G11 & 800  & 1600  & +1, --1  & toroidal & 564   \\
		G14 & 800  & 4694  & +1       & planar   & 3064  \\
		G18 & 800  & 4694  & +1, --1  & planar   & 992   \\
		G22 & 2000 & 19990 & +1       & random   & 13359 \\
		G34 & 2000 & 4000  & +1, --1  & toroidal & 1384  \\
		G38 & 2000 & 11779 & +1       & planar   & 7688  \\
		G39 & 2000 & 11778 & +1, --1  & planar   & 2408  \\
		G47 & 1000 & 9990  & +1       & random   & 6657  \\
		\hline
	\end{tabular}
\end{table*}

All evaluations are performed on G-set MaxCut benchmark instances summarized in
\cref{tab:gset_benchmarks}.
For each instance, the cut value obtained at the end of annealing is normalized by the
best-known solution for that instance.
Reported performance values correspond to the mean normalized cut value averaged across all
G-set instances.

Unless otherwise stated, all results use a fixed total simulation time of 500~ns.
This ensures that performance differences reflect architectural trade-offs rather than unequal
annealing durations.

\subsection*{Reproducibility}

\begin{table*}[t]
	\centering
	\caption{Key parameters used in the simulations. Ranges indicate values swept in the evaluation.
		Unless otherwise noted, parameters are fixed across all experiments.}
	\label{tab:simulation_parameters}
	\begin{tabular}{lll}
		\hline
		Symbol & Meaning & Value / Range \\
		\hline
		$N$ & Number of logical spins &
		$\{800, 1000, 2000\}$ (G1/G6/G11/G14/G18/G22/G34/G38/G39/G47) \\
		$\tau$ & Update interval &
		$\{2.5, 5, 7.5, 10, 15, 20\}$ ns \\
		$d$ & Hardware delay & 5 ns \\
		$c$ & Time-multiplexing reuse factor &
		$\{1, 1.25, 1.5, 2, 3, 5, 10, 15, 20, 30\}$ \\
		$b$ & DAC resolution (bits) &
		$\{1, 2, 3, 4, 6, 8, 10, 12\}$ \\
		$b_{\mathrm{ref}}$ & Reference DAC resolution & 12 \\
		\hline
	\end{tabular}
\end{table*}

The full set of simulation parameters and sweep ranges is summarized in
\cref{tab:simulation_parameters}.
All simulations are conducted using identical annealing schedules and delay assumptions across
update policies to ensure a fair comparison.
The simulation code used to generate the results is publicly available, enabling full
reproducibility of the reported experiments.

\section*{Conclusion}

We have developed a unified performance--cost landscape for parallel p-bit Ising machines
by jointly analyzing update synchronization, hardware delay, time-multiplexed p-bit reuse,
and input-DAC precision under realistic hardware constraints.
Through systematic simulations on benchmark Ising and QUBO problems, we demonstrated that
the stability and efficiency of parallel p-bit annealing are governed not only by algorithmic
update rules, but critically by the interaction between update statistics and hardware timing.

A central finding of this work is that synchronous architectures are not inherently prone to
instability, as often assumed, but can achieve stable and efficient annealing dynamics when
update simultaneity is properly controlled.
By introducing time-multiplexed reuse of physical p-bits and structured synchronous control
policies, we showed that the effective update rate can be reduced without altering the target
stationary distribution, thereby decoupling statistical correctness from physical resource count.

This decoupling enables the number of physical p-bits and input DACs to scale approximately
as $1/c$, allowing synchronous architectures to access low-cost operating regimes that are
fundamentally inaccessible to fully asynchronous updates.
In contrast, asynchronous architectures are structurally constrained by hardware delay,
requiring slower operation to maintain stability and preventing coordinated p-bit reuse.

We further demonstrated that input-DAC precision is a flexible design parameter rather than
a strict requirement.
Across a wide range of benchmark instances, near-optimal performance can be achieved using
low-resolution input DACs (typically 3--4 bits) when annealing time is appropriately adjusted,
enabling additional reductions in hardware area and power.

Taken together, these results establish time-multiplexed reuse combined with structured
synchronous control as a key architectural principle for scalable probabilistic computing
hardware.
The unified performance--cost framework presented here provides concrete design guidelines
for balancing solution quality, hardware efficiency, and timing constraints, and offers a
systematic foundation for the development of large-scale, energy-efficient p-bit-based
optimization accelerators.

\section*{Data availability}

The Python simulation code used to generate the results in this study is publicly available
in a GitHub repository~\cite{Onizawa2026parallel_pbit}.

	\section*{Funding}

This research was supported in part by KIOXIA Corporation
and by a research grant from the Murata Science and
Education Foundation. 

\section*{Author contributions statement}
N. O. conducted and analyzed the experiments. T. H. discussed the experiment. All authors reviewed the manuscript. 

\section*{Additional information}
Competing financial interests: The authors declare no competing financial interests.

\end{document}